**Added value of dynamical downscaling in sub-seasonal tropical cyclone forecast**

[a]Taehyung Kim, [a]Eunji Kim, [b]Haerin Park, [a,c]Dong-Hyun Cha*, [a]nd [d]Johan Lee

[a] Department of Civil, Urban, Earth, and Environmental Engineering, Ulsan National Institute of Science and Technology, Ulsan 44919, Republic of Korea

[b] Department of Atmospheric Sciences, Yonsei University, Yonsei-Ro 50, Seoul 03722, Korea

[c] Research & Management Center for Particulate Matters at the Southeast Region of Korea, Ulsan National Institute of Science and Technology, Ulsan 44919, Republic of Korea

[d] Climate Research Department, National Institute of Meteorological Sciences, Seogwipo, Republic of Korea

*Correspondence to: Dong-Hyun Cha, Department of Civil, Urban, Earth, and Environmental Engineering, Ulsan National Institute of Science and Technology, Ulsan 44919, Republic of Korea. Email: dhcha@unist.ac.kr

**Abstract**

Improving the sub-seasonal forecast of tropical cyclones (TCs) remains a significant challenge for climate models. This study evaluated the characteristics of tropical cyclones (TCs) in sub-seasonal forecasts using the Global Seasonal Forecast System 6 (GloSea6), an operational seasonal-to-sub-seasonal forecasting model operated by the Korea Meteorological Administration (KMA) during June–September (JJAS) from 1993 to 2016 over the western North Pacific (WNP). GloSea6 was found to underestimate TC frequency, tracks, particularly in mid-latitudes, lifetime, and intensity across all months, with the most significant errors occurring in August. To address these deficiencies, we examined whether applying dynamical downscaling to GloSea6 during August 2016, a period characterized by the lowest TC forecast skill in GloSea6, could improve sub-seasonal TC forecasts. The application of dynamical downscaling demonstrated added value by directly improving the simulation of TCs in terms of frequency, structure, intensity, and lifetime, although slight overestimations were observed. Furthermore, improved reproductions of the Indian monsoon and circumglobal teleconnection (CGT), both of which strongly influence the western North Pacific subtropical high (WNPSH), contributed to more accurate forecasts of WNPSH variability and, consequently, better predictions of TC activity in the mid-latitudes and East Asia. Therefore, dynamical downscaling can significantly advance sub-seasonal TC forecasts by not only directly improving TC characteristics but also indirectly enhancing the environmental fields (e.g., WNPSH, Indian monsoon, and CGT) that are critical to TC activity.

## 1. Introduction

Tropical Cyclones (TCs) are extremely destructive natural disasters, causing extensive socioeconomic damage through loss of human life and property (Mendelsohn et al., 2012; Daniell et al., 2013; Weinkle et al., 2018; McAneney et al., 2019; Pielke, 2021).

Their impact is especially severe in East Asia, including China, Japan, and Korea (Miller et al., 2008; Park et al., 2015; Wang et al., 2016; Li et al., 2017; Choi et al., 2019). Several studies have suggested that the damage caused by TCs in East Asia, including the Korean Peninsula, is likely to intensify due to climate change and global warming effects (Mendelsohn et al., 2012; Stocker et al., 2014). There has been a notable increase in TC activity over East Asia in recent years. For example, in 2019, seven TCs affected the Korean Peninsula, more than twice the climatological average (Shimozono et al., 2020; Kim et al., 2021; Min et al., 2021; Moon and Ha, 2021; Xu et al., 2022; Zhou et al., 2022). This highlights the urgent need for a highly accurate prediction system for TC activity to mitigate TC-related damage effectively.

Significant efforts have been made to study tropical cyclones; nevertheless, accurately predicting them remains a formidable challenge for climate models. The Sub-seasonal to Seasonal Prediction (S2S) project, jointly initiated by the World Weather Research Program and World Climate Research Program under the World Meteorological Organization (Vitart et al.,2017; Vitart and Robertson, 2018), has provided a foundation for numerous studies exploring the sub-seasonal predictability of TCs (Camargo et al., 2019; Camp et al., 2015, 2019, 2020, 2024; Garcia-Franco et al., 2023, 2025, Hansen et al., 2025; Kim et al., 2023; Klotzbach et al., 2019; Lee et al., 2018; Lee et al., 2020; Loi et al., 2025; Robertson et al., 2020; Schreck Ⅲ et al., 2023; Vitart and Robertson, 2018; Vitart and Stockdale, 2001; Viatart et al., 2025).

Previous studies showed the predictability of sub-seasonal TCs using forecasting models participating in the Sub-seasonal to Seasonal (S2S), but the models' forecast skill remains insufficient. While forecast skill has been improved with advancements in the models, predictability is still unreliable when the forecast lead time exceeds 7 days (Camp et al., 2018; Lee et al., 2018; Vitart and Robertson, 2018; Camargo et al., 2019; Kim et al., 2023). Despite these limitations, improving the sub-seasonal predictability of TC is essential, particularly for applications in climate-sensitive sectors. Accurate sub-seasonal TC forecasts can provide critical lead time for decision-making in renewable energy management, agriculture, food supply, water resources, coastal resources, fisheries, tourism and disaster risk reduction. Enhanced predictability during the peak TC season can thus contribute to more effective preparedness, and socio-economic resilience in vulnerable regions.

Climate models, such as models that participated in the S2S project with coarse resolution struggle to accurately capture fine-scale information such as orography, and atmospheric dynamics, limiting their effectiveness in predicting extreme weather events (Xu et al., 2019, 2021). Dynamical downscaling, a method that addresses this limitation, utilizes limited-area, high-resolution regional climate models (RCMs) driven by initial and boundary conditions from a global climate model (GCM), providing more detailed weather and climate information (Giorgi and Mearns, 1991; Giorgi et al., 2009; Meng et al., 2011; Xu et al., 2019). This method, grounded in physical principles, is widely applied in regional climate predictions and future climate projections (Cha et al., 2016a, 2016b; Jin et al., 2016; Juzbašić et al., 2024; Kim et al., 2025; Kim et al., 2018, 2020, 2021; Kim et al., 2023; Lee et al., 2017, 2023; Lee et al., 2013; Lee and Cha, 2020; Lee et al., 2020; Park et al., 2016, 2021a, 2021b, 2022, 2024; Tang et al., 2016; Shin et al., 2024; Xu et al., 2019, 2021). However, whether dynamical downscaling can improve the sub-seasonal prediction of TCs, and to what extent, remains uncertain.This improvement largely depends on how well the model captures key large-scale circulation features that govern TC variability.The western North Pacific subtropical high (WNPSH) is a key circulation system affecting the western North Pacific TC activities (Ho et al., 2005; Kim et al.,2011; Wang et al., 2013; Choi et al., 2015; He et al., 2015; Choi et al., 2017; Wang and Wang, 2019; Chen et al., 2022). Depending on the location and extension of the WNPSH, it could impact mid-latitude TC activity. The strong convection over northwestern India is known to trigger anomalous upper-troposphere highs over South Korea during summer through the Rossby wave propagation (Kim et al., 2019, 2020; Yeo et al., 2019; Min et al., 2020). An unusually strong Indian monsoon may have contributed to the exceptionally high number of TCs affecting East Asia (Min et al., 2021). In August 2016, an anomalously high number of TCs occurred, most of which affected the mid-latitudes including East Asia. A similar teleconnection pattern aforementioned was observed during this period. Also, the ridge of the WNPSH was located to the north on 30N in such a way that it was readily affected by the mid-latitude system in this period. The WNPSH plays a crucial key role in simulating the mid-latitude activity of TCs, and the WNPSH forecast skill may also be linked to the reproduction of Indian monsoon and circumglobal teleconnection (CGT).

In this study, we investigated the characteristics of sub-seasonal TC forecasting with GloSea6, the global seasonal forecasting system of the Korea Meteorological Administration (KMA). Furthermore, this study focuses on August 2016 to assess whether dynamical downscaling improves TC forecast skill and to evaluate the added value (AV) it produces. Section 2 describes the model, data, methods, and design of the dynamical downscaling and numerical experiment used in the study. Section 3 presents the sub-seasonal TC forecast skill of GloSea6, along with the dynamical downscaling results and their added values (AVs) for August 2016. The summary and conclusions are provided in Section 4.

## 2. Models, datasets, and methods

### 2.1 Models

Global Seasonal Forecast system version 6, GloSea6-GC3.2 (Kim et al., 2021), is used in this study, which is one of the state-of-the-art fully coupled dynamical prediction systems to generate S2S forecast at the KMA. It consists of the following components:

- Atmosphere: Unified Model version 11.5 (UM; Davis et al. 2005), Global Atmosphere 7.2 (Williams et al. 2015; Walters et al. 2017)
- Land surface: Joint UK Land Environment Simulator version 5.6 (JULES; Best et al. 2011; Clark et al. 2011), Global Land 8.0 (Williams et al. 2015; Walters et al. 2017)
- Ocean: Nucleus for European Modeling of the Ocean version 3.4 (NEMO; Madec 2008), Global Ocean 6.0
- Sea-ice: The Los Alamos Sea Ice Model version 5.1.2 (CICE; Hunke et al. 2010), Global Sea-Ice 8.1
- Coupler: Ocean Atmosphere Sea Ice Soil3 (OASIS3; Valcke 2013)

GloSea6 hindcast data were utilized for the June to September (JJAS) over a 24-year span from 1993-2016. The performance of GloSea6, which includes 84 ensemble members per year, is evaluated using operational hindcasts initialized from four start dates (May, June, July, and August). These hindcasts are initialized on three consecutive weeks (00UTC on the 9th, 17th, and 25th of each month) centered around the middle day of the month preceding the analysis period (e.g., using the 9th, 17th, and 25th of July for analyzing August). This evaluation focuses on verifying TC forecasting at a sub-seasonal time scale (15–45 days). Each initial date had seven ensemble members (001~007) using the stochastic physics scheme Stochastic Kinetic Energy Backscatter v2 (SKEB2; Bowler et al. 2009). This methodology is the same as Kim et al. (2023), but there is a difference in ensemble members on each initial date. The horizontal resolution and vertical levels in GloSea6 are the same as in its previous version, GloSea5, with the atmospheric model consisting of 85 vertical levels at a horizontal resolution of about 60km at mid-latitudes (N216; 0.83° longitude × 0.55° latitude) and ocean model consisting of 75 vertical levels at a horizontal resolution of 0.25° (ORCA025; ~27 km on the equator).

The Advanced Research Weather Research and Forecast (WRF) model version 4.4.0 (Skamarock et al., 2019) was used to verify the impact of dynamical downscaling in sub-seasonal TC forecast (Table 1). The model consisted of one fixed domain with 20km horizontal resolutions. The model made use of the MSKF (Zheng et al., 2016) cumulus parameterization scheme, the WSM6 (Hong and Lim, 2006) cloud microphysics scheme, the YSU planetary boundary scheme (Hong et al., 2006; Noh et al., 2003), and the CAM radiation scheme (Collins et al., 2004). We used the GloSea6 Hindcast data, which was initialized on July 25th, 2016, with a temporal resolution of 6 hours and a horizontal spatial resolution of 0.83° × 0.56° as the initial conditions and the boundary condition of the WRF model. We conducted 1-month simulations (00UTC Aug 1st, 2016, to 00UTC Sept 1st, 2016) for 7 ensemble members.

The Convective Available Potential Energy (CAPE) time scale, which effectively reduces atmospheric instability at coarse grid resolutions, is proportional to the grid resolution DX. However, higher model grid resolution increases the resolved cloud area and accelerates the saturation process, leading to the rapid dissipation of CAPE within the CAPE time scale, which can cause strong convection. To mitigate these issues, the MSKF scheme incorporates an adjustment time scale based on Bechtold et al. (2008), scaled by a parameter $\beta$ influenced by the horizontal grid scale (Zheng et al., 2016). $\beta$ is defined as:

$$\beta = 1 + \ln\left(\frac{25}{\mathrm{DX}}\right) \quad (1)$$

, where β is 1.0 (2.8) at the 25-km (4-km) model grid spacing in Equation 1. In the enhanced convection experiment, grid-scale convection was increased by reducing the scale-aware parameter (β) to 0.5, which amplified the effects of the cumulus parameterization scheme (CPS). Conversely, in the weakened convection experiment, grid-scale convection was reduced by increasing the scale-aware parameter (β) to 3.0 (Figure 4.5). The modified parameters were specifically applied to domain 2 to artificially strengthen or weaken the Indian monsoon. All other experimental configurations remained consistent with WRF experiment settings (Table 1).

**2.2 Observations, reanalysis**

The sub-seasonal TC forecast skill in GloSea6 and WRF was verified using the US Navy Joint Typhoon Warning Center Best Track database (JTWC), including the center of the TC (latitude and longitude), maximum sustained wind speed (Chu et al. 2002). Here, we stipulated a TC as a system with a 1-min maximum sustained wind speed of 34 knots (17.5 m/s) or higher. To analyze environmental fields that affect TC activity and intensification, hourly atmospheric variables were obtained from the ECMWF Reanalysis 5th generation (ERA5) dataset (Hersbach et al., 2020) with horizontal resolution of 0.25°. These variables include geopotential height (GPH) at 200 and 500 hPa, mid-level vertical velocity (500hPa omega), specific humidity at 700 hPa, wind and relative vorticity at 850 hPa, and latent heat flux (LHF).

**2.3 Method**

The domain for TC analysis was set as 0°N–50°N and 90°E–180°E, representing the area where TC activity is highest over the WNP. The analysis domain of the environmental field related to TC activity and CGT was set to 0°N-55°N and 60°E-180°E, which is an extension of the TC analysis domain. To identify the large-scale teleconnection pattern associated with WNPSH, GPH and wave activity flux (WAF; Takaya and Nakamura, 2001) at 200 hPa were used.

The detection and tracking methods for TCs simulated by GloSea6 and WRF were adapted from previous studies (Kim et al., 2023). The detection and tracking algorithms used were as follows: (1) identification of potential tropical storms based on local minimum sea-level pressure; (2) Application of a search radius of 500 km centered on the minimum sea-level pressure; 3) Confirmation of maximum sustained surface wind exceeding a threshold of 16 m $s^{-1}$ within the search radius; (4) Requirement for maximum relative vorticity at 850 hPa to exceed $4.5 \times 10^{-5} s^{-1}$; (5) Verification of a warm core temperature exceeding 1.5 K over three vertical pressure levels (700, 500, and 300hPa); (6) Validation that maximum wind speed at 850 hPa is greater than that at 300 hPa; (7) Establishment of a minimum lifetime criterion of 2 days for all storms with a warm core.; (8) Tracking of TC paths originating from the identified potential storms. TC genesis and track density were computed by binning the storm genesis location and track locations into 5° × 5° grid boxes. The genesis and track density were normalized by dividing them by the ensemble size and length of the analysis year.

## 3. Results

### 3.1 Characteristics of Sub-seasonal Tropical Cyclone Prediction in GloSea6

The monthly mean of the TC genesis density in the GloSea6 and JTWC best track are shown in Fig. 1. GloSea6 reasonably reproduced the locations of TC genesis each month, along with their expansion from the South China Sea (SCS) to the International Date Line as the season progressed. However, GloSea6 underestimated TC genesis density in the South China Sea (SCS) for all months and underrated it in the east of the Philippines except for July. In particular, August, which is the most active TC month, showed the largest difference in TC genesis density and frequency (Table 2). Fig. 2 shows the monthly mean TC track density from the JTWC best track data and GloSea6. Similar to the genesis density results, GloSea6 consistently underestimated the overall track density compared to the best track data, with the largest discrepancy observed in August. In particular, mid-latitude TC activity (above 30°N) was under-simulated from July to September.

Annual time series of TC frequency for each month from 1993 to 2016 are shown in Fig. 3. The TC frequency difference in June per year simulated by GloSea6 was the smallest, and interannual variability (Temporal correlation: 0.55) was properly simulated. In July, there was an error in TC frequency in 1993, 1994, and 1996, but otherwise, the GloSea6 reproduced TC frequency and interannual variability (temporal correlation: 0.29) reasonably, although not as well as in June. In contrast, in August and September, GloSea6 had large interannual TC frequency differences, underestimated TC frequency in most years, and failed to simulate interannual variability (Temporal correlation in August: 0.14, in September: -0.14).

The averaged monthly TC frequency, lifetime, and intensity are shown in Table 1. As with genesis density, the TC frequency of GloSea6 had the most significant error in August (-1.3), followed by September (-1.0), July (-0.6), and June (-0.4). TC lifetime error for GloSea6 was largest in June (-2.2 days), followed by August (-0.9 days), July (-0.5 days), and September (-0.1 days). For TC intensity in GloSea6, the error is largest in September (-20.0ms$^{-1}$), and is smaller in the order of June (-19.3ms$^{-1}$), August (-17.9ms$^{-1}$), and July (-17.8ms$^{-1}$). To summarize sub-seasonal TC prediction skill in GloSea6, it underestimated TC frequency, TC track, especially in mid-latitude activities, TC lifetime, and intensity in all months, particularly the errors are large in August.

3.2 Impact of Dynamical Downscaling in Sub-seasonal TC Forecast and its Direct Added Value

GloSea6 exhibited the lowest tropical cyclone (TC) forecast skill in August, particularly in 2016 (Fig. 3c). This study aims to assess whether applying dynamical downscaling during August 2016, a period marked by the lowest TC forecast accuracy in GloSea6, could improve sub-seasonal TC forecasts.

The results of JTWC best track, GloSea6, and WRF simulation in averaged TC frequency, TC lifetime, and lifetime maximum intensity (LMI) for August 2016 are presented in Fig. 4. During this period, 9 TCs were observed, exceeding the climatological mean. GloSea6 underestimated the TC frequency, predicting only 5 TCs, whereas WRF improved upon this with a simulation of 9.14 TCs, closely aligning with observation (Fig. 4a). GloSea6 also underestimated the average TC lifetime, predicting 3.8 days compared to observed 4.6 days, while WRF overestimated it at 6.0 days (Fig. 4b). The observed average LMI of TCs was 32.5 ms$^{-1}$, whereas GloSea6 predicted a weaker intensity of 21.9 ms$^{-1}$. WRF slightly overestimated the LMI at 40.7 ms$^{-1}$ but still represented an improvement (Fig. 4c). Although WRF tended to overestimate both TC intensity and lifetime relative to observations, it represented a significant improvement over the underestimation by GloSea6.

Fig. 5 represents the simulated azimuthally averaged radial and tangential wind fields for all TCs at the time of LMI. In the following analysis, these results are examined to assess how dynamical downscaling influences the horizontal and vertical structural characteristics of TC simulations. GloSea6 tended to reproduce comparatively weaker tangential winds, and decrease low-level inflow and upper-level outflow (Fig. 5b). This result indicated that GloSea6 underestimated both the inner core structure of TCs and weaker primary and secondary circulation. In contrast, WRF more realistically captured the inner core structure, and these circulations, though it tended to slightly overestimate the overall TC structure (Fig. 5c). In summary, the application of dynamical downscaling led to clear added value by improving TC frequency, intensity, lifetime, and structure (both primary and secondary circulation) directly, although it represented somewhat overestimation.

Near-surface low-pressure circulation and low-level convergence enhance the release of LHF, which intensifies upward motion and drives both the intensification and maintenance of TCs. This physical process acts as a positive feedback loop, progressively strengthening the TC . Therefore, if this physical intensification process and its associated positive feedback are well represented in the model, the intensity of TC can be realistically reproduced. Drawing on this framework, we analyzed why GloSea6 underestimated TC intensity and examined how the WRF generated added value in simulating TC intensity (Fig. 6). These results were area-averaged within a 300km radius from the TC center. The x-axis represents the TC timestep as a percentage of the TC lifetime across the entire ensemble of TCs. GloSea6 reproduced weakerlow-level convergence than reanalysis data and a rapid weakening of convergence over the TC lifetime, while WRF showed improved forecast skill in convergence at 850-hPa (Fig. 6a). LHF, which serves as the primary energy source for TC activity, gradually decreased throughout the TC lifecycle in ERA5. GloSea6 reasonably captured this trend, while WRF also reproduced the decrease in LHF similarly to ERA5 but slightly overestimated its magnitude (Fig 6b). Omega at 500 hPa, representing vertical motion at mid-level atmosphere, was somewhat overestimated in both GloSea6 and WRF (Fig 6c). Observations showed that TC maximum wind speed (MWS) generally increases and then decreases over the TC lifecycle. GloSea6 not only failed to reproduce this pattern but also significantly underestimated MWS. In comparison, WRF accurately captured both the temporal variation and magnitude of MWS, although it slightly was overestimated (Fig 6d). Several environmental factors can impede or delay the tropical cyclone development process, but among them, the most significant are vertical wind shear (VWS) and dry air in the mid-troposphere. VWS, which inhibits TC intensification, showed an increasing trend over the TC lifecycle in both models. However, GloSea6 exaggerated VWS, producing unfavorable conditions for TC intensification. WRF simulated VWS similarly to ERA5 by representing it as weaker than GloSea6. The moisture in the mid-level atmosphere, another factor inhibiting TC intensification, decreased with the TC lifecycle in ERA5, and both models reproduced this characteristic adequately (Fig. 6f).

In summary, GloSea6 failed to capture the positive feedback mechanism of TC intensification but exaggerated VWS, leading to unfavorable conditions for sustained TC activity and premature TC dissipation. Consequently, GloSea6 produced shorter-lived and weaker TCs. In contrast, WRF demonstrated improved forecast skill in TC intensity, accurately capturing the positive feedback mechanisms driving TC intensification and better representing environmental variables linked to TC suppression, in line with reanalysis data. As a result, WRF simulated stronger and longer-lived TCs.

### 3.3 Indirect Added Value in sub-seasonal TC forecast by the dynamical downscaling

To assess whether the dynamical downscaling improved the simulation performance of environmental fields associated with TC activities, we analyzed the 10-day averages of 850 hPa wind, relative vorticity at 850 hPa, and 500 hPa geopotential height (Fig. 7).

In August 2016, WNPSH, located east of Japan, strengthened until mid-August before weakening toward the end of the month. During this time, southwesterly winds extended unusually far eastward, generating a monsoon trough (MT) centered around 15~20°N. This MT stretched from the Indochina Peninsula to the west of the international dateline, with peak activity near 155°E, where southwesterly and easterly flows converged (Fig. 7a, b, c). In early August, GloSea6 simulated a weaker WNPSH, which then rapidly contracted in mid-August. This flaw led to a weakening of the MT. Since more than 80% of the TCs in the WNP are driven by monsoon circulation, GloSea6's inadequate simulation of the monsoon trough resulted in an underestimation of TC genesis (Fig. 7d, e, f). In early August, WRF closely simulated WNPSH and MT compared to the ERA5. After mid-Aug, although the simulated WNPSH became somewhat distorted, its overall reproduction, including MT, remained reasonably accurate. By late August, however, WRF misplaced the WNPSH too far to the south (Fig. 7g, h, i). Still, dynamical downscaling proved effective for improving the simulation of environmental fields associated with tropical cyclone (TC) genesis and activity. Although the errors are smaller than those in GloSea6, they tend to grow over longer simulation periods. Therefore, we investigated how WRF enhanced the performance of the WNPSH simulation.

During summer, intense convection over northwestern India is recognized for triggering anomalous upper-troposphere highs over East Asia, driven by the Rossby wave propagation. In August 2016, the WNPSH was persistently and strongly positioned east of Japan for an extended period (Fig. 7a, b, c). Concurrently, strong convection, indicated by outgoing longwave radiation (OLR) anomalies (Fig. 8a), occurred over northwestern India. This was accompanied by a distinct 200 hPa geopotential height anomaly and wave activity flux, indicating Rossby wave propagation associated with the CGT. The intensified WNPSH development was associated with anomalous anti-cyclonic circulation in the upper troposphere over the northeast of Japan (Fig. 8a). GloSea6 underestimated the strength of the Indian monsoon and failed to reproduce the westward Rossby wave propagation (Fig. 8b). In contrast, WRF reasonably simulated the Indian monsoon and CGT, showing improvements over GloSea6's results (Fig. 8c).

In August 2016, a total of nine TCs occurred, one in the South China Sea (SCS, ~122°E) and eight in the Northern Pacific (NP, over 122°E). Of these, seven TCs affected East Asia (Over 30°N and ~150°E). This is influenced by the earlier analyzed position of WNPSH, which guided TCs toward mid-latitudes. GloSea6 underrated the number of TCs affecting mid-latitudes (2.8 TCs) and East Asia (2 TCs), as well as the total number of TCs (5 TCs) and TC genesis in the NP (4.3 TCs). However, it relatively accurately simulated TC genesis in the SCS (0.7 TCs). These inaccuracies stemmed from poor simulation skills of the WNPSH, MT, and CGT in GloSea6. In contrast, WRF, with its improved simulation of WNPSH, MT, and CGT, demonstrated enhanced forecast skill for TC genesis (total: 9.1 TCs; SCS: 1.4 TCs; NP: 7.7 TCs) and TC activity in mid-latitudes (5.3 TCs) and East Asia (3.4 TCs), which GloSea6 underestimated (Fig. 9).

To investigate the simulation performance relationships between CGT, Indian monsoon, and the WNPSH in WRF, enhanced and weakened convection experiments (hereafter ENC and WKC, respectively) over northwestern India were conducted. The environmental fields reproduced by ENC and WKC are shown in Fig. 10. In ENC, WNPSH and MT were stronger and maintained their positions until mid-August, with the WNPSH expanding westward in late August (Fig. 10a). The WKC, on the other hand, simulated similarly to ERA5 in early August. However, by mid-August, the westerly winds from the Indian Ocean weakened due to the rapid westward expansion of the WNPSH (Fig. 10b). By comparing the results of these two experiments, we found that the strengthening of the Indian monsoon in WRF influences the strength, location, and expansion of the simulated WNPSH.

Additionally, we analyzed whether this convection-enhanced (-weakened) experiment affected the strengthening of CGT (Fig. 11). ENC simulated wave activity flux (WAF) similarly to WRF but simulated stronger upper-tropospheric high over the WNP (Fig. 11a). This suggests that WRF can reproduce the strengthening of upper troposphere anti-cyclonic circulation caused by enhanced zonal Rossby wave propagation, which subsequently leads to a strengthening of the WNPSH and delays its westward extension over the forecast lead time. Conversely, WKC simulated a weaker upper level high over the WNP as well as WAF (Fig. 11b). The weakening of the upper troposphere high, driven by reduced zonal Rossby wave propagation, accelerated the westward extension of the WNPSH over the forecast lead time. These findings emphasize that the accuracy of the Indian monsoon simulation in WRF not only affects the performance of WNPSH reproduction but also significantly impacts the simulation of CGT. In the ENC experiment, the maintenance and westward expansion of the Western North Pacific Subtropical High (WNPSH) were further delayed, whereas in the WKC experiment, the expansion of the WNPSH was accelerated. We investigated whether TC activity in mid-latitudes changed with WNPSH forecast skill (Fig. 12). The results showed an increase in TC activities over mid-latitude and East Asia in the ENC experiment, while a decrease was observed in the WKC experiment. Notably, despite an increase in TC genesis over the NP in WKC, TC activity over East Asia decreased significantly.

In summary, improvements in reproducing in the Indian monsoon and CGT pattern led to better forecasts of the WNPSH. This, in turn, resulted in improved forecasts of TC activity, particularly in mid-latitude and East Asia. While WRF did not generate direct added value on TC simulation, it indirectly enhanced TC activity forecasts by improving the simulation of related environmental factors, such as the CGT, Indian monsoon, and WNPSH.

## 4. Summary and conclusion

This study evaluated the sub-seasonal prediction skill of the GloSea6 model for tropical cyclones (TCs) and assessed the impact of dynamical downscaling using the WRF model.

GloSea6 consistently underestimated TC genesis density in the SCS for all months and also underestimated it east of the Philippines, except in July. The largest discrepancy in TC genesis density and frequency (-1.3) was particularly pronounced in August, the most active TC month. Similar to the genesis density result, the overall track density was consistently underestimated in GloSea6 compared to observation data, with the largest difference presented in August. Additionally, mid-latitude TC activity (above 30°N) was under-simulated from July to September. Interannual variability of TC in GloSea6 was reasonably simulated in June (0.55) and July (0.29), but poorly simulated in August (0.14) and September (-0.14). The TC lifetime errors in GloSea6 were most significant in June (-2.2 days), followed by August (-0.9 days), July (-0.5 days), and September (-0.1 days). Regarding TC intensity in GloSea6, the largest error occurred in September (-20.0$ms^{-1}$), followed by progressively smaller errors in June (-19.3$ms^{-1}$), August (-17.9$ms^{-1}$), and July (-17.8$ms^{-1}$). In summary, the sub-seasonal TC prediction skill of GloSea6 consistently underestimated TC frequency, track accuracy (especially in mid-latitude regions), lifetime, and intensity across all months, with particularly significant errors observed in August.

We applied dynamical downscaling to August 2016, which presented the lowest sub-seasonal TC forecast skill in GloSea6, to verify whether it improved sub-seasonal TC predictability. During August 2016, a total of nine TCs occurred, which is anomalously high compared to normal. GloSea6 underestimated the TC frequency, predicting only five TCs, whereas WRF improved on TC frequency by reproducing 9.14 TCs. Where GloSea6 simulated shorter TCs lifetime and weaker intensity, WRF reproduced the improved results, although with slight overestimations. TC structure and primary and secondary circulations were also improved by applying dynamical downscaling, which is underestimated in GloSea6 but also slightly over-simulated. To analyze TC intensity predictability, we employed the CISK mechanism, which is one of the possible mechanisms of positive feedback in TC intensification. GloSea6 not only failed to properly simulate the positive feedback of TC intensification but also exaggerated vertical wind shear (VWS), which hindered TC activity and led to rapid dissipation of TCs. Consequently, GloSea6 reproduced a shorter lifetime and weaker TCs. In contrast, WRF improved forecast skill in TC intensity by successfully simulating the positive feedback from TC intensification and accurately capturing environmental variables that suppress TC intensification, similar to reanalysis data. This allowed WRF to simulate stronger and longer-lived TCs.

Environmental fields such as wind at 850 hPa and 500 GPH related to TC activity were analyzed. The WNPSH based on GPH at 500 hPa, located east of Japan, intensified through mid-August but weakened toward the end of the month. GloSea6 simulated a weaker WNPSH in early August, which rapidly contracted in mid-August. This deficiency led to a weakening of the monsoon trough. Because of this failure to accurately simulate the monsoonal trough, there was an underestimation of TC genesis. WRF, on the other hand, closely simulated the WNPSH and MT in early August compared to ERA5 data. However, after mid-August, WRF reproduced a distorted WNPSH but still reasonably captured the overall pattern of the WNPSH and MT. Dynamical downscaling demonstrated improvements in simulating the environmental fields associated with TC genesis and activity. In August 2016, the WNPSH was located persistently and strongly east of Japan for an extended period, while strong convection occurred over northwestern India, accompanied by a clear CGT. The intensified development of the WNPSH was associated with anomalous anti-cyclonic circulation developed in the upper troposphere over the northeast of Japan. GloSea6 underestimated the Indian monsoon and failed to reproduce the westward Rossby wave propagation, whereas WRF reasonably simulated the Indian monsoon and CGT, improving on the results reproduced by GloSea6. The high number of mid-latitude and East Asia impact TCs during this period was due to the strong Indian monsoon and Rossby wave propagation, which located the WNPSH as a guide for TCs to move toward mid-latitude. GloSea6, due to its poor simulation of the WNPSH, MT, and CGT, underrated TC activity in mid-latitudes and East Asia. In contrast, WRF improved the forecast skill for TC activity in mid-latitudes and East Asia through its better simulation of the WNPSH, MT, and CGT. Experiments with artificially increasing (decreasing) convection over northwestern India were conducted to verify the simulation performance relationship between CGT, Indian monsoon, and WNPSH in WRF. The ENC further delayed the maintenance and expansion of the WNPSH, producing a favorable environment for TCs to move into mid-latitude and East Asia, while the WKC accelerated the expansion of the WNPSH, generating an unfavorable environment for TCs to move into the mid-latitude and East Asia. The results showed an increase in mid-latitude and East Asia TC activities in the ENC and a decrease in mid-latitude and East Asia TC activities in the WKC. Improving the simulation of the Indian monsoon and CGT can lead to improvement in TC activity and related environmental fields, highlighting the importance of large-scale teleconnections in bridging the weather and climate, particularly on the sub-seasonal time scale.

In this study, we found added value in applying dynamical downscaling that directly improved the sub-seasonal forecast of the TC in terms of frequency, intensity, and structure. Dynamical downscaling also indirectly generated added value in TC activity by improving the predictability of Indian monsoon, CGT, and WNPSH. It is also meaningful that this research contributes to improving the forecast skill of S2S models. However, advanced techniques, such as coupled air-sea interaction are necessary to improve sub-seasonal TC forecasts using dynamical downscaling. Furthermore, this study represents a singular case study on the sub-seasonal prediction of TC over the western North Pacific. More cases and periods should be explored to evaluate whether and to what extent dynamical downscaling can improve the sub-seasonal TC forecasts.

## Acknowledgment

This work was funded by the Korea Meteorological Administration Research and Development Program under Grant KMI (RS-2023-00241809). This work was supported by a grant from the National Institute of Environment Research (NIER), funded by the Ministry of Environment (MOE) of the Republic of Korea (NIER-2021-03-03-007).

# Tables

**Table 1** List of Experiments and their configuration conducted in this study

| Experiment | WRF | ENC | WKC |
|---|---|---|---|
| Model Version | | WRF Model Ver. 4.4.0 | |
| Initial Date | | 2016. 07. 25 | |
| Simulated Period | | 2016. 08 | |
| Ensemble member | 7 members (001-007) | 3 members (001-003) | 3 members (001-003) |
| Projection Resolution | 20km (D01) | 20km (D01)<br>10km (D02) | 20km (D01)<br>10km (D02) |
| Convective Parameterization Scheme | MSKF | MSKF<br>Same as WRF but for a different scale aware parameter in D02 : 0.5 | MSKF<br>Same as WRF but for a different scale aware parameter in D02 : 3.0 |
| Cloud Micro Physics Scheme | | WSM6 | |
| Planetary Boundary Layer | | YSU | |
| Longwave Radiation Scheme | | CAM | |
| Shortwave Radiation Scheme | | CAM | |
| Land Surface Model | | Unified NOAH LSM | |

**Table 2** Tropical Cyclone (TC) Frequency, Duration, and Intensity averaged per each month for climatology (1993-2016) in the JTWC best track data (BST), GloSea6 (GLS6), and differences between the two.

| | Frequency (number of TC) | | | | Duration (days) | | | | Intensity ($ms^{-1}$) | | | |
|---|---|---|---|---|---|---|---|---|---|---|---|---|
| | June | July | Aug | Sept | June | July | Aug | Sept | June | July | Aug | Sept |
| BST | 1.6 | 3.7 | 5.2 | 5.1 | 5.8 | 5.3 | 5.6 | 4.8 | 40.4 | 40.5 | 40.0 | 42.9 |
| GLS6 | 1.2 | 3.1 | 3.9 | 4.1 | 3.6 | 4.8 | 4.7 | 4.7 | 21.1 | 22.7 | 22.1 | 22.9 |
| Diff. | -0.4 | -0.6 | -1.3 | -1.0 | -2.2 | -0.5 | -0.9 | -0.1 | -19.3 | -17.8 | -17.9 | -20.0 |

# Figures

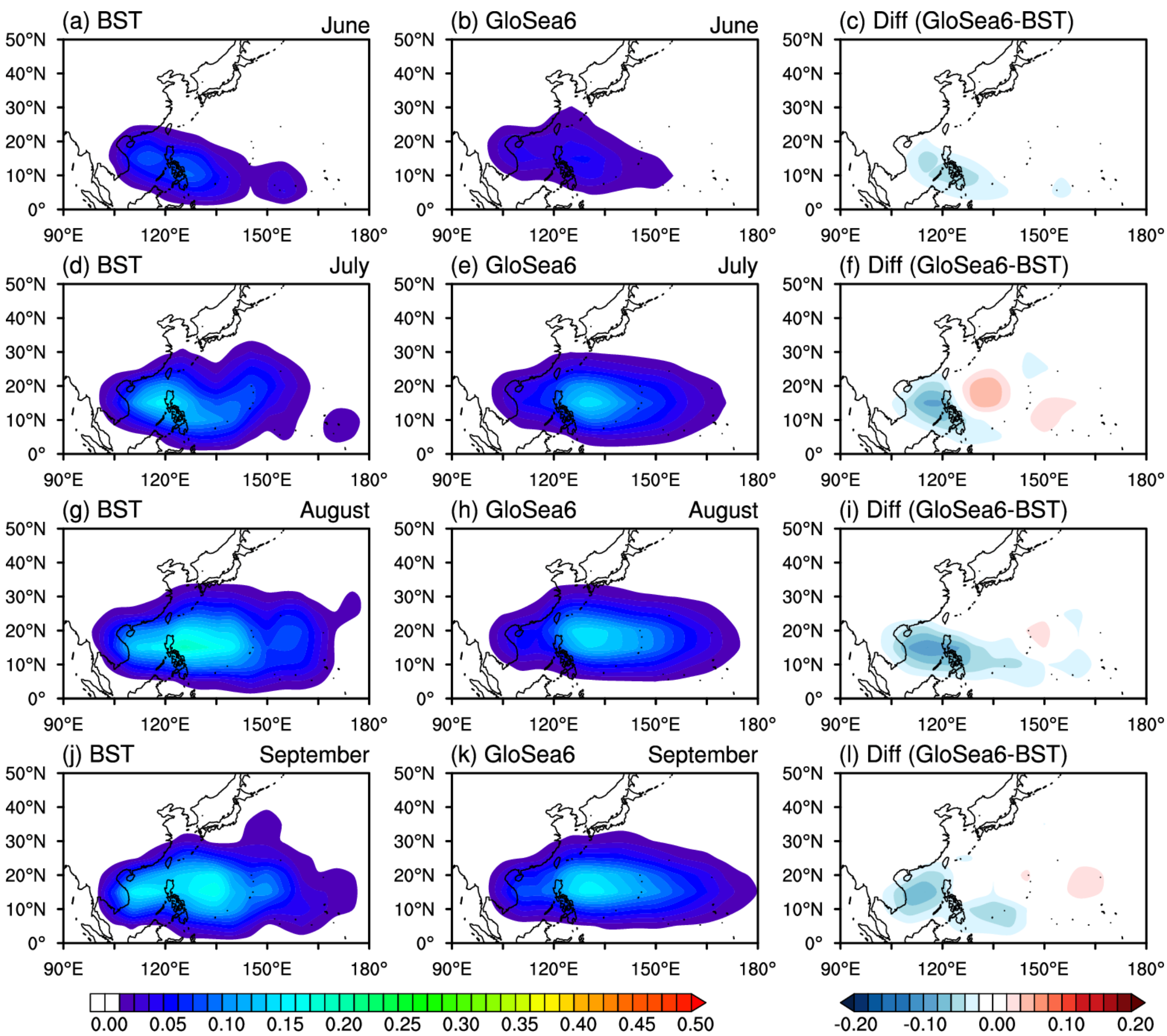

Figure 1. Composite maps of the monthly climatological mean tropical cylcone (TC) genesis density for June (first row), July (second row), August (third row), and September (forth row), averaged over the 24-year period from 1993 to 2016. Results are shown for (a, d, g, j) the JTWC best track data, (b, e, h, k) GloSea6, and (c, f, i, l) the difference between GloSea6 and JTWC best track data.

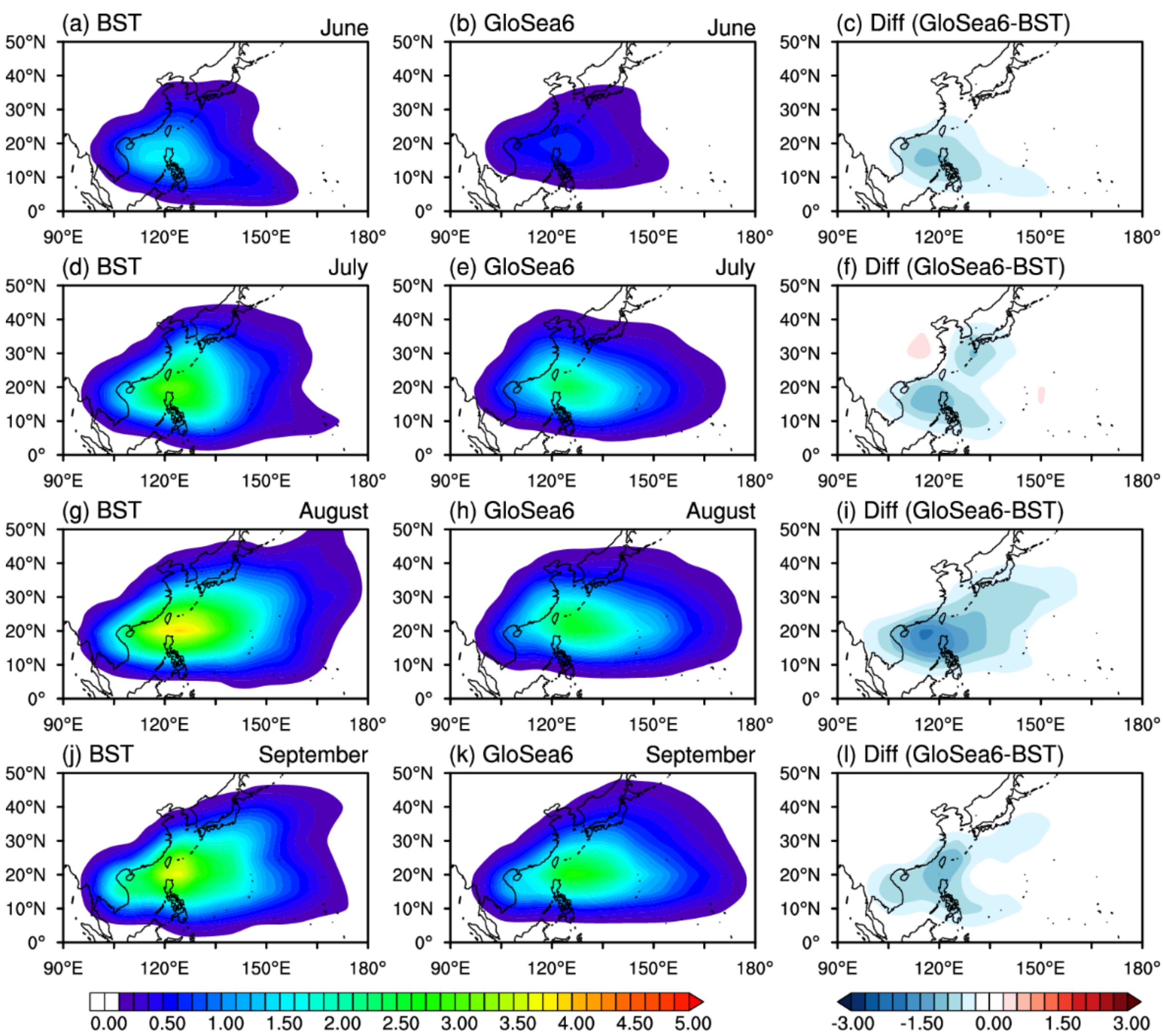

Figure 2. Same as Figure 1 but for composite map of TC track density.

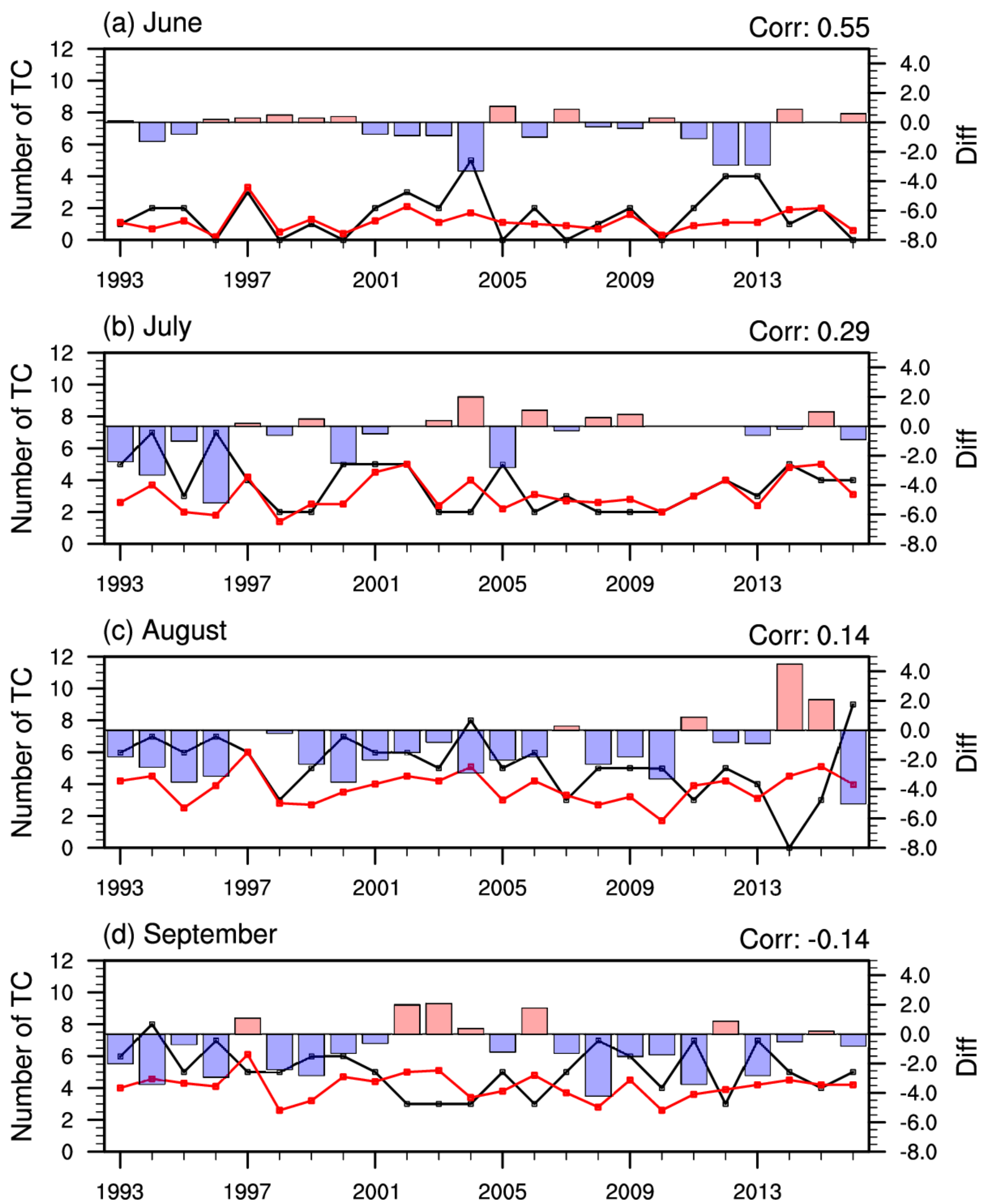

Figure 3. Annual time series of TC frequency for each month during 1993-2016. The black dotted line denotes JTWC best track data, the red dotted line denotes GloSea6, with their temporal correlation shown in the upper-right corner. Red bars indicate positive differences, and blue bars indicate negative differences.

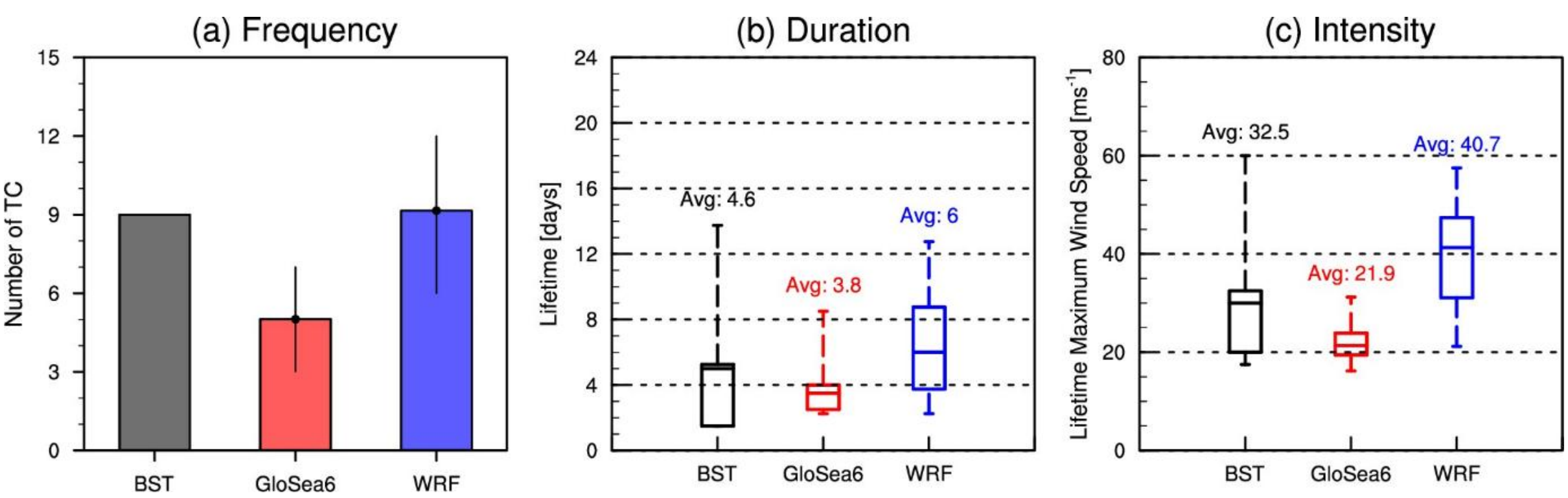

Figure 4. (a) TC frequency, (b) TC duration, and (c) intensity in Aug 2016, respectively. Black box denotes JTWC best track, red box denotes GloSea6, and blue box denotes WRF.

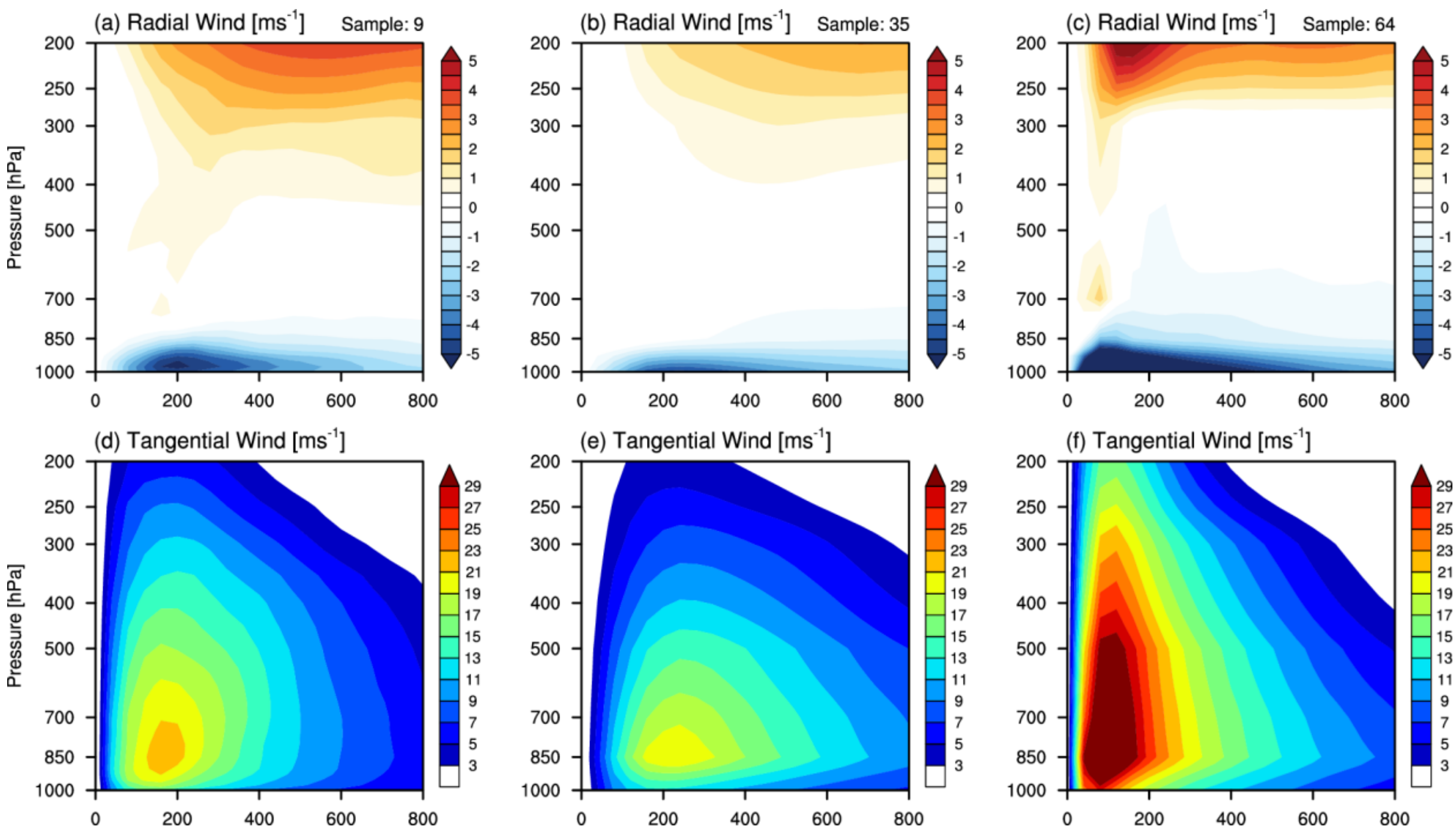

Figure 5. Azimuthally averaged vertical-radial structure of TC: (a, b, c) radial wind ($m \cdot s^{-1}$, shading) and (d, e, f) tangential wind ($m \cdot s^{-1}$, shading) at lifetime maximum intensity in Aug 2016 from ERA5 reanalysis (left), GloSea6 (middle), and WRF (right). The number of ensemble indicates top-right corner.

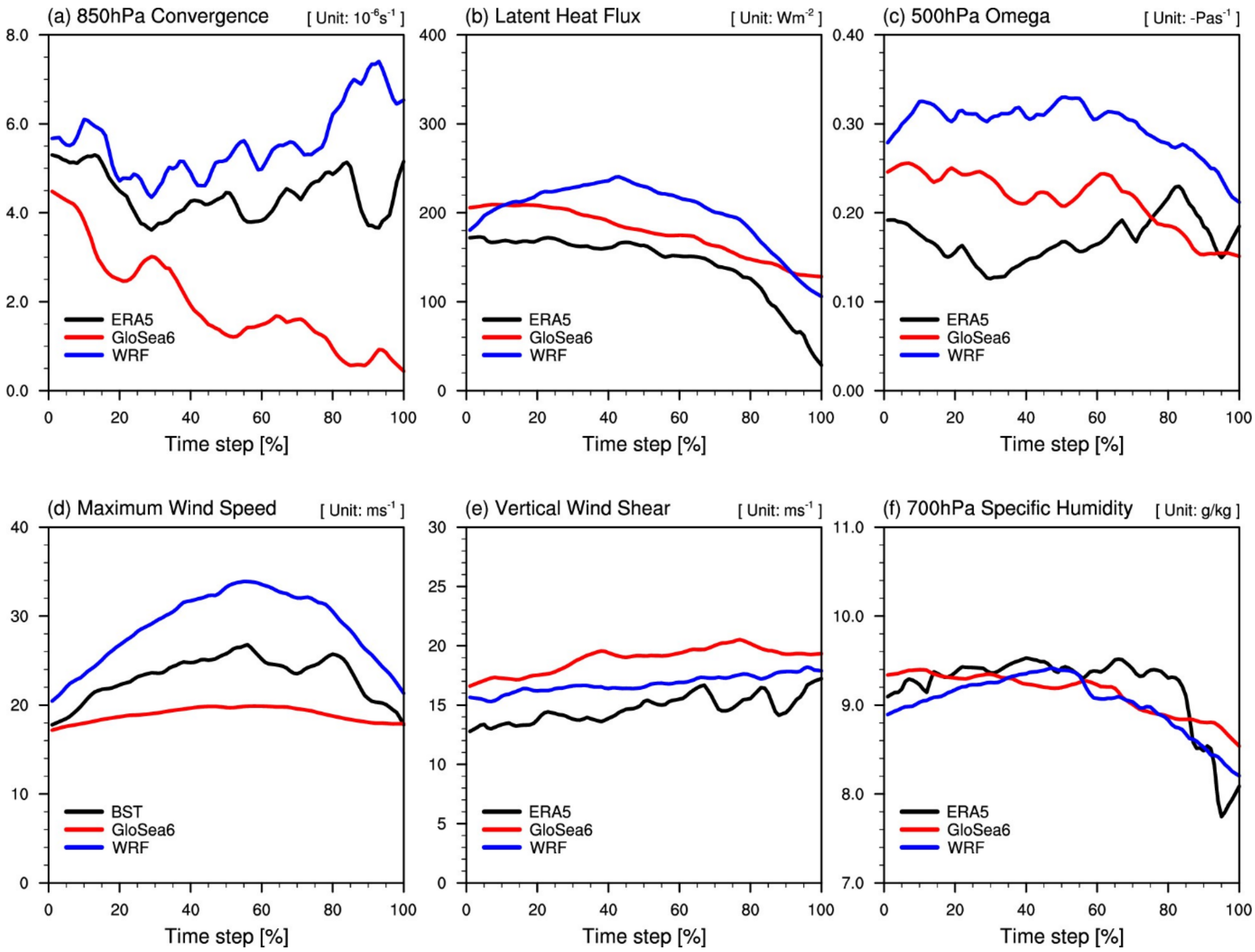

Figure 6. Time series of area-averaged variables within a 300km radius from TC center during its lifetime: (a) Convergence at 850 hPa ($10^{-6}s^{-1}$), (b) Latent heat flux ($W \cdot m^{-2}$), (c) Omega at 500 hPa ($-Pa \cdot s^{-1}$), (d) Maximum wind speed ($m \cdot s^{-1}$), (e) Vertical wind shear ($m \cdot s^{-1}$), and (f) Specific Humidity at 700 hPa ($g \cdot kg^{-1}$). Black line indicates JTWC best track data and ERA5 reanalysis data, red line denotes GloSea6 result, and blue line indicated WRF result, respectively.

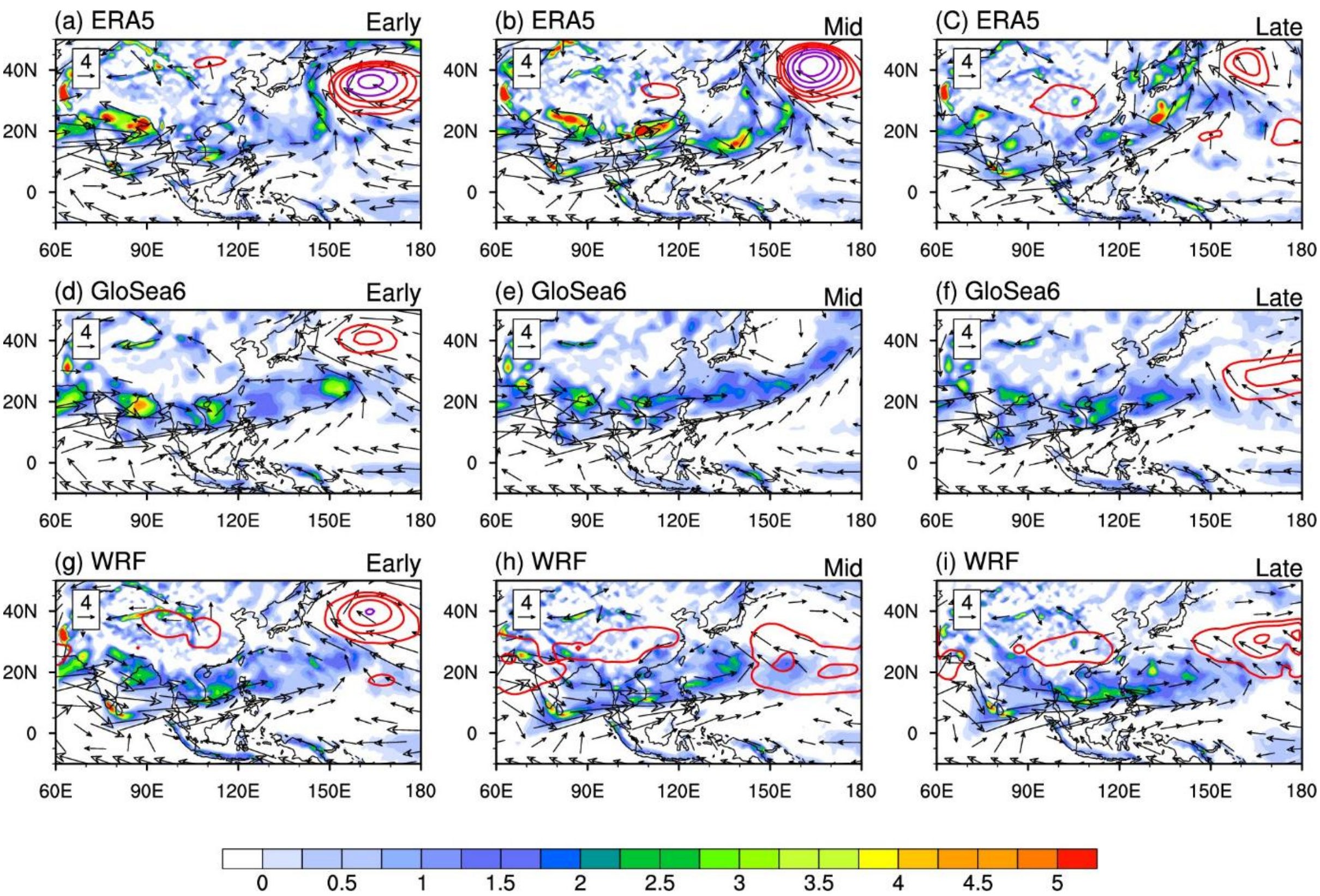

Figure 7. Composite maps of 850 hPa wind above 4m·s$^{-1}$ (vector; m·s$^{-1}$), 850 hPa relative vorticity (shaded; 10$^{-5}$s$^{-1}$), and 500 hPa geopotential height above 5890m (contour; m) distribution in early (left; a, d, g), mid (mid; b, e, h), and late (right; c, f, i) August 2016 corresponding to (a, b, c) ERA5 data, (d, e, f) GloSea6, and (g, h, i) WRF

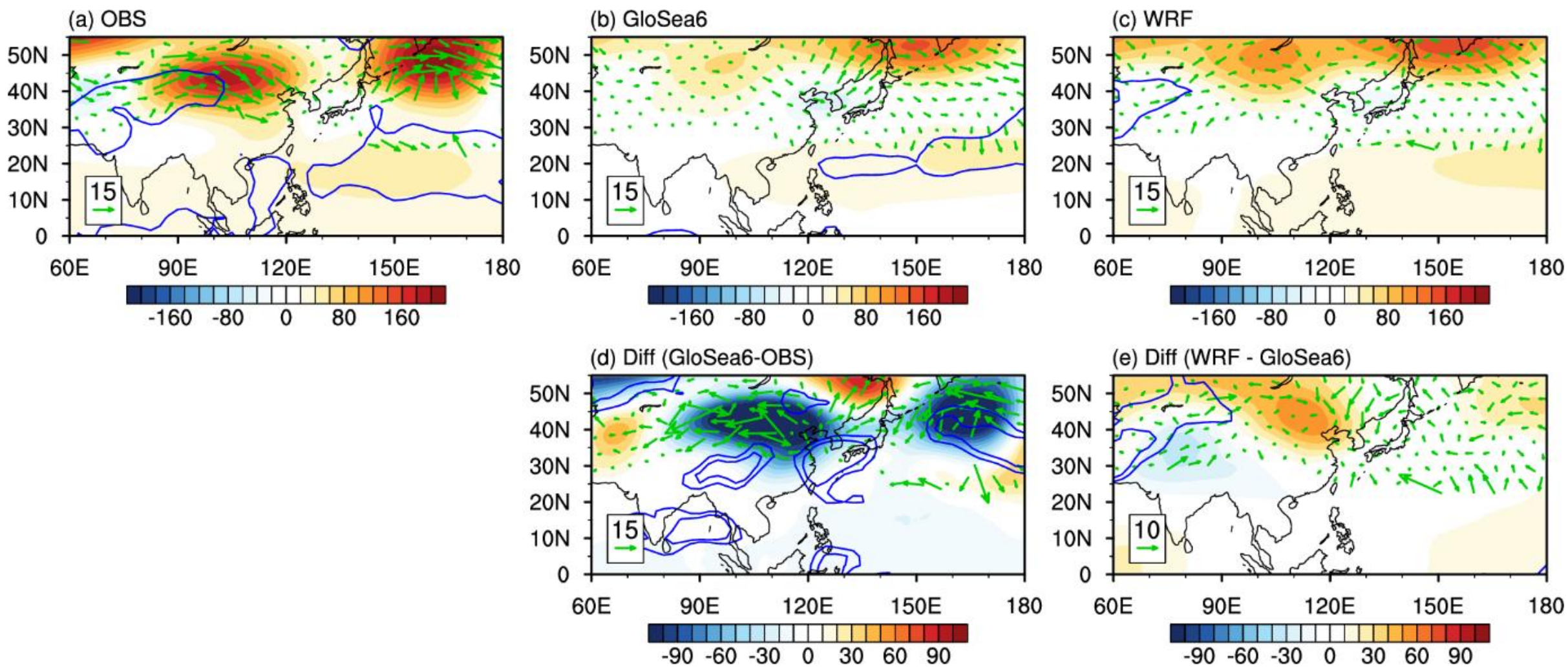

Figure 8. Composite maps of 200 hPa wave activity flux (green vector; $m^2 \cdot s^{-2}$), 200 hPa geopotential height anomaly (shaded; m), and outgoing longwave radiation (contour; $Wm^{-2,}$ below -10 $Wm^{-2}$) distribution in Aug 2016 corresponding to (a) ERA5 data (200 hPa GPH, wave activity flux) and NOAA OLR data, (b) GloSea6, and (c) WRF, (d) the difference between GloSea6 and ERA5, and (e) the difference between WRF and GloSea6

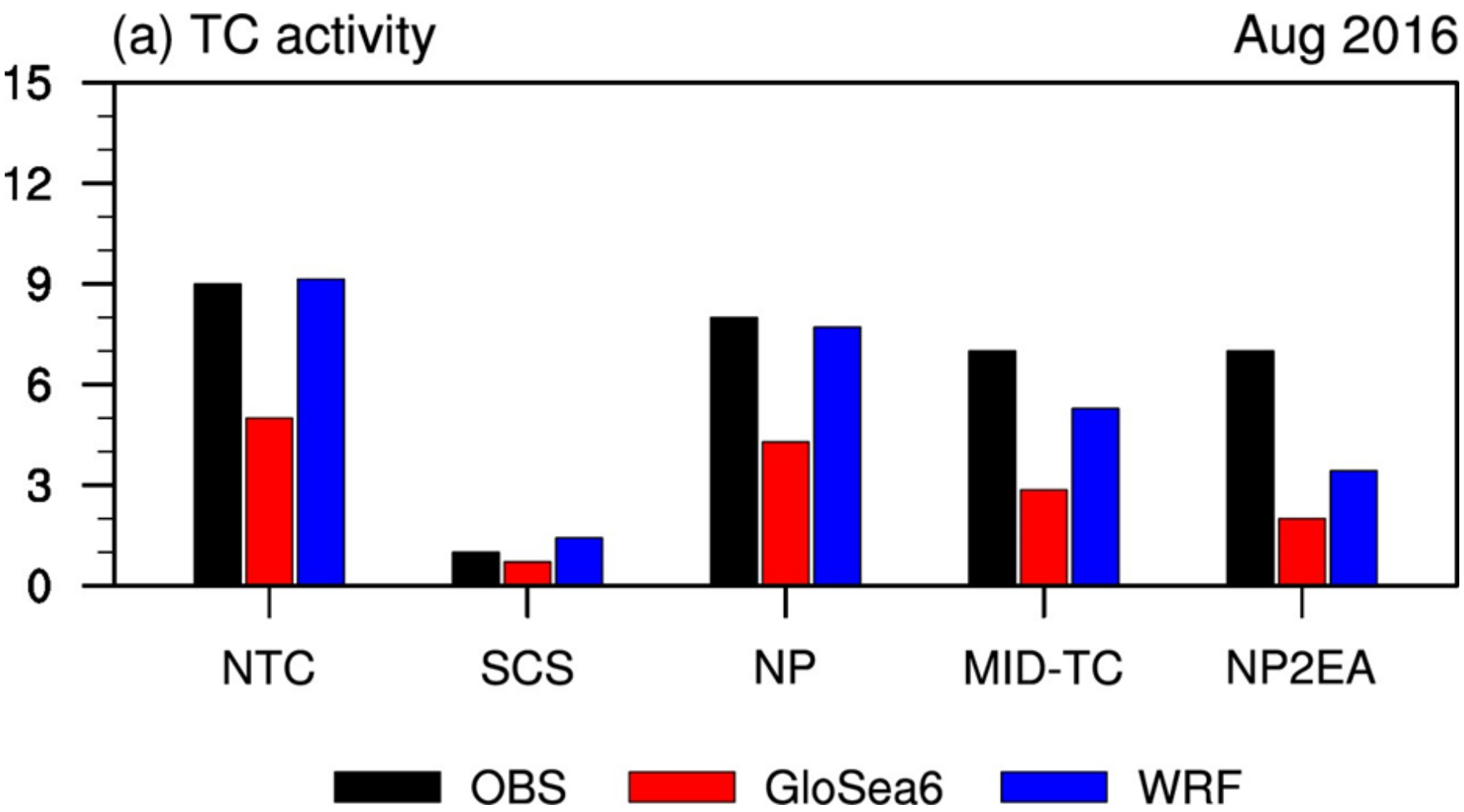

Figure 9. The number of total TC (NTC), genesis in the South China Sea (SCS), genesis in the Northern Pacific (NP), mid-latitude (above 30N) activity TC (MID-TC), and far East Asia activity TC which generated in the NP (NP2EA) in Aug 2016. Black box denotes JTWC best track data, red box denotes GloSea6 result, and blue box denotes WRF results, respectively.

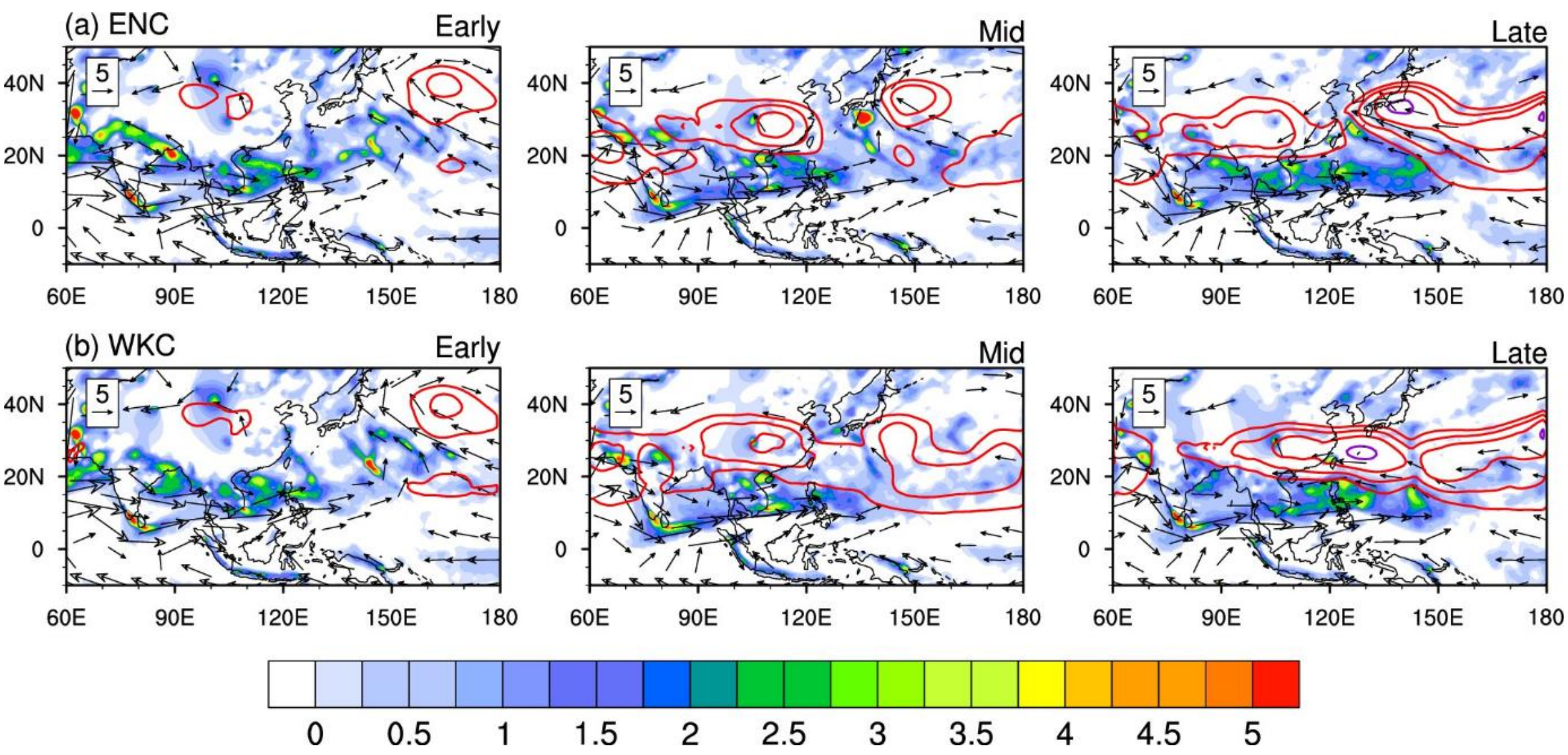

Figure 10. Same as Figure 8 but for composite map of (a) ENC experiment result, and (b) WKC experiment result

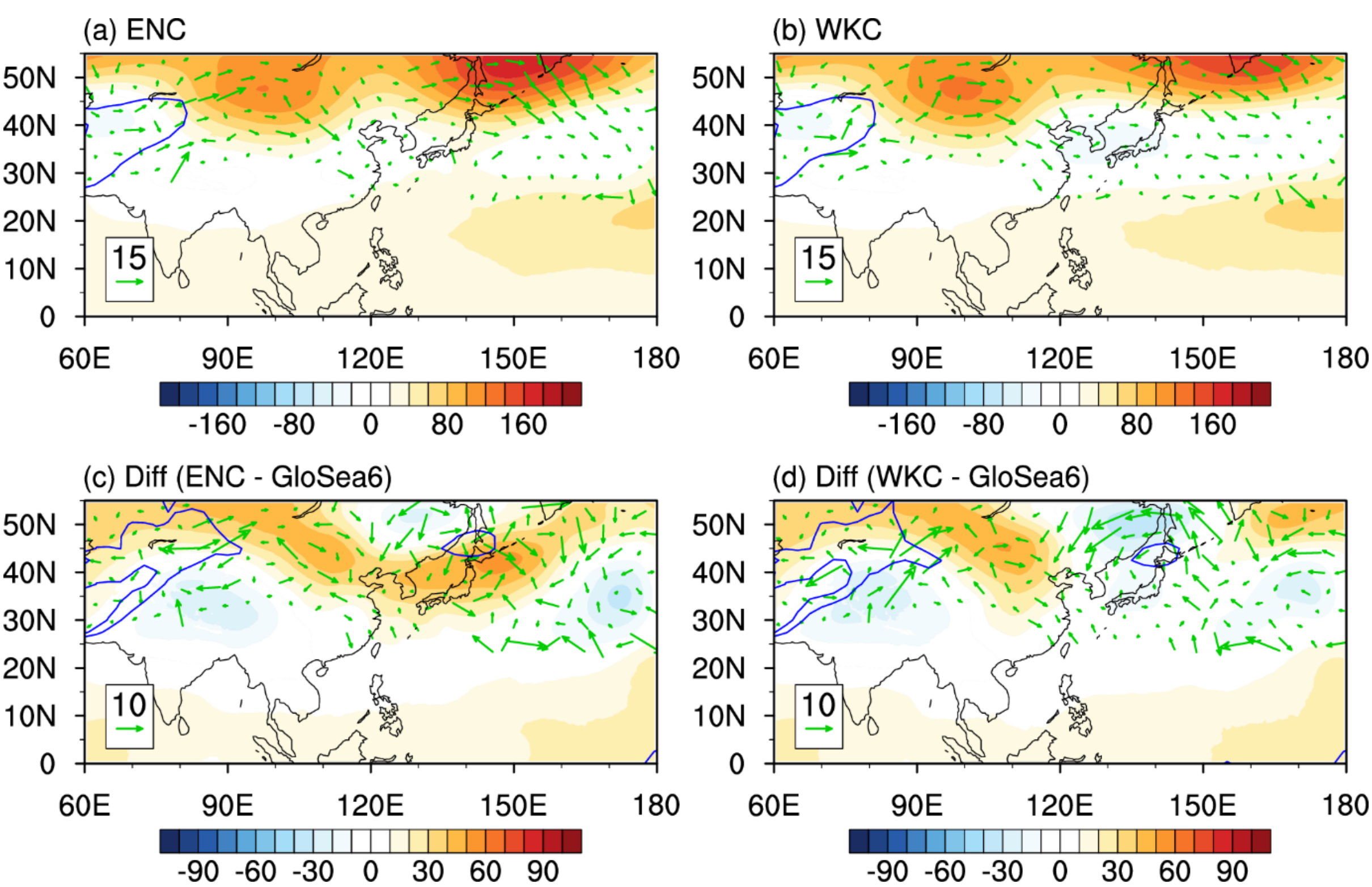

Figure 11. Same as Figure 8, but for composite map of (a) ENC, (b) WKC, and difference between GloSea6 and (c) ENC and (d) WKC.

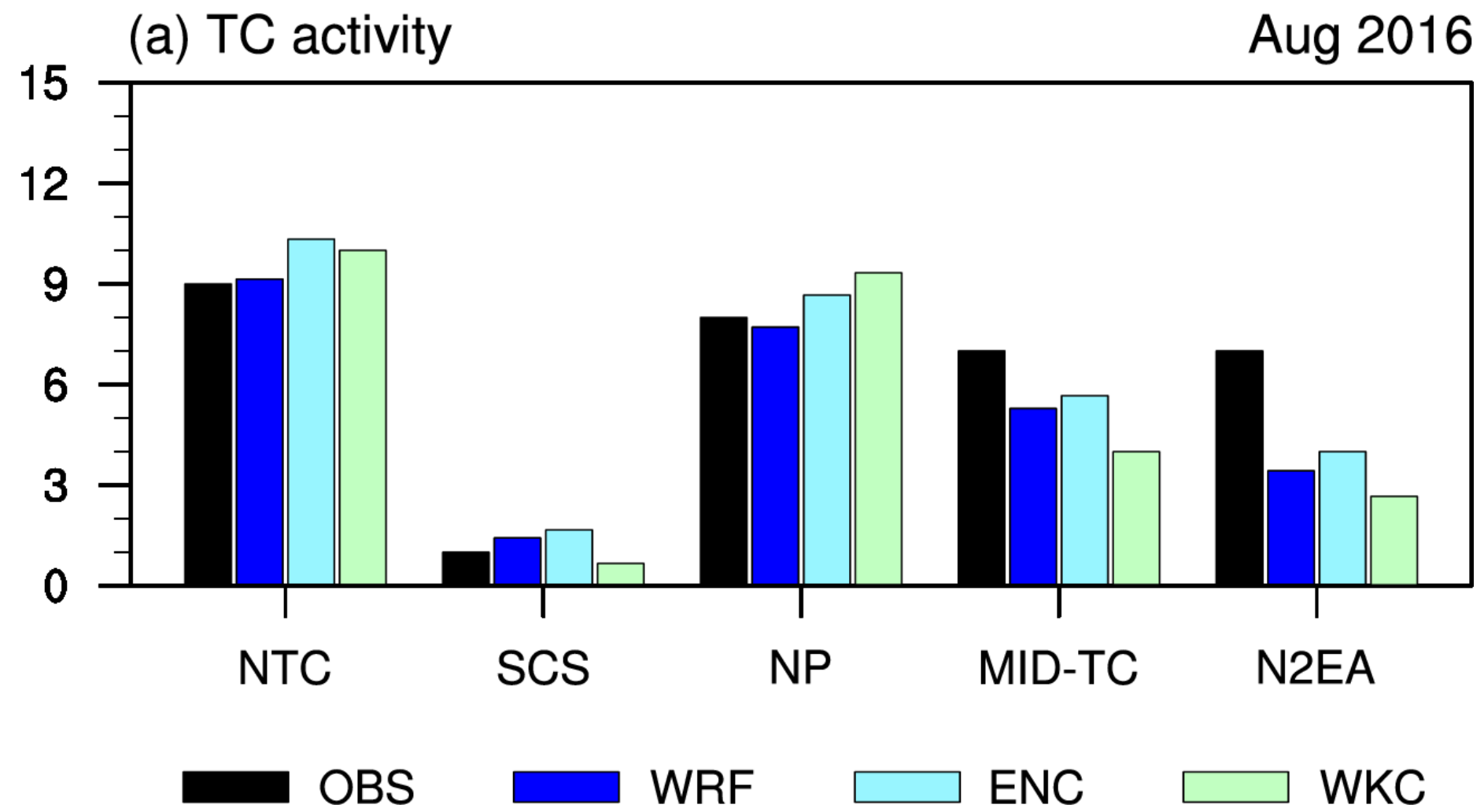

Figure 12. Same as the Figure 9 but for ENC (light blue) and WKC (light green)